\documentclass[aps,reprint,floatfix,nofootinbib]{revtex4-1}
\usepackage{amsmath,amssymb,graphicx}
\usepackage{hhline,bbm}
\usepackage{float}
\usepackage{subcaption}

\usepackage[T1]{fontenc}

\makeatletter
\DeclareRobustCommand{\cev}[1]{%
  \mathpalette\do@cev{#1}%
}
\newcommand{\do@cev}[2]{%
  \fix@cev{#1}{+}%
  \reflectbox{$\m@th#1\vec{\reflectbox{$\fix@cev{#1}{-}\m@th#1#2\fix@cev{#1}{+}$}}$}%
  \fix@cev{#1}{-}%
}
\newcommand{\fix@cev}[2]{%
  \ifx#1\displaystyle
  \mkern#23mu
  \else
  \ifx#1\textstyle
  \mkern#23mu
  \else
  \ifx#1\scriptstyle
  \mkern#22mu
  \else
  \mkern#22mu
  \fi
  \fi
  \fi
}

\makeatother

\newcounter{theoremcounter}
\stepcounter{theoremcounter}

\newtheorem{definition}{Definition}

\newcommand{\bs}{\boldsymbol}
\newcommand{\bra}[1]{\left\langle #1\right|}
\newcommand{\ket}[1]{\left|#1\right\rangle}

\newcommand{\ketbra}[2]{\ket{#1}\bra{#2}}

\makeatletter
\newcommand{\vast}{\bBigg@{4}}
\newcommand{\Vast}{\bBigg@{5}}
\makeatother

\begin{document}

\title{The Non-Disjoint Ontic States of the Grassmann Ontological Model, Transformation Contextuality, and the Single Qubit Stabilizer Subtheory}
\author{Lucas Kocia}
\affiliation{National Institute of Standards and Technology, Gaithersburg, Maryland, 20899, U.S.A.}
\author{Peter Love}
\affiliation{Department of Physics, Tufts University, Medford, Massachusetts, 02155, U.S.A.}
\begin{abstract}
  We show that it is possible to construct a preparation non-contextual ontological model that does not exhibit ``transformation contextuality'' for single qubits in the stabilizer subtheory. In particular, we consider the ``blowtorch'' map and show that it does not exhibit transformation contextuality under the Grassmann Wigner-Weyl-Moyal (WWM) qubit formalism. Furthermore, the transformation in this formalism can be fully expressed at order \(\hbar^0\) and so does not qualify as a candidate quantum phenomenon. In particular, we find that the Grassmann WWM formalism at order \(\hbar^0\) corresponds to an ontological model governed by an additional set of constraints arising from the relations defining the Grassmann algebra. Due to this additional set of constraints, the allowed probability distributions in this model do not form a single convex set when expressed in terms of disjoint ontic states and so cannot be mapped to models whose states form a single convex set over disjoint ontic states. However, expressing the Grassmann WWM ontological model in terms of non-disjoint ontic states corresponding to the monomials of the Grassmann algebra results in a single convex set. We further show that a recent result by Lillystone \emph{et al}. that proves a broad class of preparation and measurement non-contextual ontological models must exhibit transformation contextuality lacks the generality to include the ontological model considered here; Lillystone \emph{et al.}'s result is appropriately limited to ontological models whose states produce a single convex set when expressed in terms of disjoint ontic states. Therefore, we prove that for the qubit stabilizer subtheory to be captured by a preparation, transformation and measurement non-contextual ontological theory, it must be expressed in terms of non-disjoint ontic states, unlike the case for the odd-dimensional single-qudit stabilizer subtheory.
\end{abstract}
\maketitle

\section{Introduction}

There has been much interest recently in the study of contextuality by those pursuing the classical simulation of near-term quantum computation. This is because of its central role in the extension of many efficiently simulatable systems to quantum universality. Contextuality has been shown to be the salient ingredient introduced in the magic state injection of Clifford circuits~\cite{Howard14,Raussendorf15}, measurement-based quantum computation~\cite{Raussendorf01,Okay17}, and the T-gate extension of the Clifford gateset~\cite{Gross06,Veitch12,Mari12,Kocia16,Kocia17_2}.

Contextuality can be present in the operational forms of preparation contextuality, transformation contextuality and measurement contextuality~\cite{Spekkens05}. Measurement contextuality is perhaps the oldest and best-known form of contextuality, and is the inability to pre-assign outcomes to a set of observables without prior knowledge of the ``context'' that they will be taken in~\cite{Redhead87,Peres90,Mermin90,Spekkens05}. In general, contextuality is believed to be a non-classical property of quantum mechanics and has been shown to require higher than order \(\hbar^0\) terms in the Wigner-Weyl-Moyal (WWM) representation of the observables~\cite{Kocia16,Kocia17_2,Kocia17_3}. It is most frequently described in the ontological models formalism, wherein measurement contextuality is responsible for multiple possible outcomes in an ontological model (defined in the next section) where a single outcome is expected~\cite{Spekkens05}.

One of the simplest quantum subtheories is the single-qubit stabilizer subtheory, which has long been thought to be completely non-contextual~\cite{Wallman12,Blasiak13,Kocia17_2}. However, recently, Lillystone \emph{et al}. proved that for a single qubit, a broad class of ontological models that are preparation and measurement non-contextual still exhibit transformation contextuality under the ``blowtorch'' map~\cite{Lillystone18}. Such a result contradicts the association of the presence of contextuality with the presence of non-classical properties.

In this paper we relate the Grassmann WWM formalism at order \(\hbar^0\) to an ontological model---a preparation and measurement non-contextual \(\psi\)-epistemic ontological model for a single qubit---and perform Lillystone \emph{et al}.'s calculations. We find that the Grassmann WWM ontological model does not exhibit transformation contextuality under the ``blowtorch'' map or any other map consisting of convex combination of one-qubit stabilizer states.

This suggests that there must be some aspect of the Grassmann WWM ontological model that is neither captured by Lillystone \emph{et al}.'s proof nor by many prior ontological models studied in the literature. In particular, we will consider the Grassmann WWM formalism in the framework of ontological models defined over non-disjoint ontic states. We find that they possess unique properties that are not captured by restricting study to ontological models defined only over disjoint ontic states, as in past studies~\cite{Spekkens05,Wallman12,Karanjai18}. The ontological model corresponding to the Grassmann WWM formalism appears to be an example of a novel subclass of ontological models that seem to have been overlooked in the literature.

We begin with a review of the results of Lillystone \emph{et al}.~\cite{Lillystone18} in Section~\ref{sec:review} where we also introduce transformation contextuality in ontological models with disjoint ontic states. In Section~\ref{sec:cptpmapinwwmformalism} we introduce the Grassmann WWM formalism and demonstrate that it does not exhibit transformation contextuality at order \(\hbar^0\). In Section~\ref{sec:example} we introduce a simple ontological model over non-disjoint states and show how re-expressing it over disjoint ontic states produces more than one convex subset. This motivates why such ontological models cannot be represented by models with disjoint states. We then demonstrate in Section~\ref{sec:grassmannHVT} that the Grassmann WWM formalism is such an ontological model with non-disjoint states and establish more of its properties in Section~\ref{sec:fundamentals}. We prove that it is inequivalent to Lillystone \emph{et al}.'s representative disjoint eight-state model in Section.~\ref{sec:inequivalence}. We conclude in Section~\ref{sec:conc}.

\section{Review}
\label{sec:review}

We define ontological models according to~\cite{Leifer14}: An ontological model is defined by a measurable space \(\Lambda\) of possible physical states,  with an associated \(\sigma\)-algebra \(\Sigma\), and sets of measures or measurable functions \(P_A:\Sigma \rightarrow [0,1]\) are used to represent preparations, transformations and measurements in the ontological model. \(\Lambda\) is called the ontic space and elements \(\lambda \in \Lambda\) are called ontic states.

An ontological model is a classical probabilty theory and so must satisfy Kolmogorov's three axioms:
\begin{enumerate}
    \item non-negativity: \(P(\lambda) \in \mathbb R \, \text{and}\, P(\lambda) \ge 0 \, \forall \, \lambda \in \Lambda\),
    \item \(P(\Lambda) = 1\),
    \item \(\sigma\)-additivitiy: \(P(\cup_i \lambda_i) = \sum_i P(\lambda_i)\) if \(\{\lambda_i\}\) are disjoint (i.e. correspond to mutually exclusive events).
\end{enumerate}
From these axioms follow~\cite{Ash08}:
for any two subsets \(A\), \(B \in \Lambda\),
\begin{enumerate}
 \setcounter{enumi}{3}
\item probability of an empty set: \(P(\emptyset) = 0\),
\item the sum rule: \(P(A \cup B) = P(A) + P(B) - P(A \cap B)\),
and,
\item monotonicity: if \(B \subset A\), then \(P(A) \le P(B)\) and \(P(A \setminus B) = P(A) - P(B)\).
\end{enumerate}

\(A\) and \(B\) are disjoint if \(P(A \cap B) = \emptyset\) and non-disjoint otherwise. It should be noted that ontological models can be defined over both disjoint and non-disjoint ontic states and past work has been careful to include both cases~\cite{Leifer14}. Non-disjoint ontological models are often treated as a ``coarse-graining'' of a disjoint ontological model. We will show that in some cases, they must be treated in terms of non-disjoint states in order that their states form a single convex set.

Furthermore, we can distinguish between two different types of ontological models. From Harrigan \emph{et al}.~\cite{Harrigan2010}:
\begin{definition}
  An ontological model is \(\psi\)-ontic if for any pair of preparation procedures, \(P_\psi\) and \(P_\phi\), associated with distinct quantum states \(\Psi\) and \(\phi\), we have \(p(\lambda|P_\psi) p(\lambda|P_\phi) = 0\) for all \(\lambda\).
\end{definition}

\begin{definition}
  \label{def:epistemic1}
  If an ontological model fails to be \(\psi\)-ontic, then it is said to be \(\psi\)-epistemic.
\end{definition}

\(\psi\)-ontic and \(\psi\)-epistemic ontological models are both also called ``hidden variable theories''. Colloquially, \(\psi\)-ontic ontological models can be thought of as hidden variable theories where the ``hidden'' variables are not really hidden (because distinct wavefunctions correspond to distinct subsets of \(\Lambda\)) while \(\psi\)-epistemic models are models with truly hidden variables.

Lillystone \emph{et al}. introduce an eight-state ontological model for one qubit~\cite{Lillystone18}, originally developed in~\cite{Wallman12}, which consists of an ontic space \(\Lambda = \{\pm 1\}^3\) that can be indexed by \(\lambda = (x,y,z) \in \Lambda_1 \) for \(x\), \(y\), \(z \in \pm 1\)---the eigenvalues of the Pauli matrices \(\hat X\), \(\hat Y\) and \(\hat Z\), respectively. This model is preparation and measurement non-contextual~\cite{Wallman12}. Ontic states evolve under the maps corresponding to \(\hat X\), \(\hat Y\) and \(\hat Z\) as
\begin{equation}
\label{eq:eightstatemap1}
  \Gamma_X: (x,y,z) \rightarrow (x,-y,-z),
\end{equation}
\begin{equation}
  \Gamma_Y: (x,y,z) \rightarrow (-x,y,-z),
\end{equation}
and
\begin{equation}
  \Gamma_Z: (x,y,z) \rightarrow (-x,-y,z),
\end{equation}
respectively.
They evolve under the Hadamard gate \(H\) as
\begin{equation}
\label{eq:eightstatemap4}
  \Gamma_H: (x,y,z) \rightarrow (z,-y,x).
\end{equation}
Since \(x\), \(y\), and \(z\) are each in \(\{\pm 1\}\), these maps are not continuous; they are permutations on \(\{\pm 1\}^3\) defined by Eqs.~\ref{eq:eightstatemap1}-\ref{eq:eightstatemap4}.

\begin{figure}[ht]
  \includegraphics{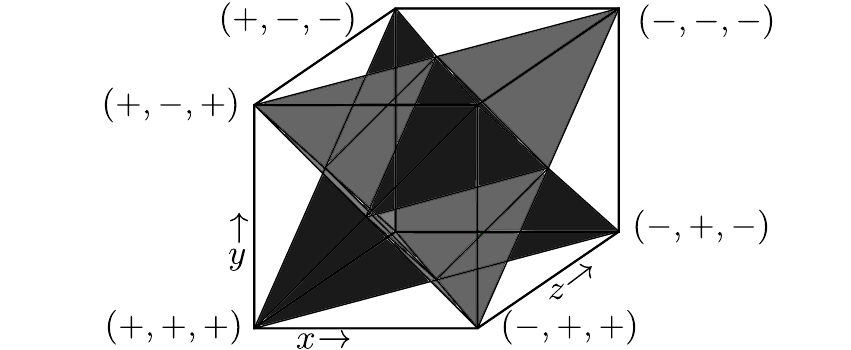}
  \caption{The eight-state ontic space reproduced from~\cite{Lillystone18} showing the simplices of the even- (dark grey) and odd- (light grey) parity ontic states. \(T_1\) maps ontic states in a tetrahedron to another ontic state in the same color tetrahedron. \(T_2\) maps ontic states between the two tetrahedral regions even though it is operationally equivalent.}
  \label{fig:8stateonticspace}
\end{figure}

Lillystone \emph{et al}. then consider evolution of an input state \(\rho\) under the two operationally equivalent implementations of the following map:
\begin{equation}
  T_1(\rho) = \frac{1}{4}(\rho + X\rho X+ Y \rho Y + Z\rho Z),
\end{equation}
and
\begin{equation}
  T_2(\rho) = H T_1(\rho) H.
\end{equation}
\(T_1(\rho) = T_2(\rho) = I/2\) and so this is often called the ``blowtorch'' map since it is akin to ``taking a blowtorch'' to the state \(\rho\) and heating it up to become the maximally mixed state (a Gibbs distribution at infinite temperature)~\cite{Brink}. Though \(T_1(\rho) = T_2(\rho) = I/2\), the authors point out that under their eight-state model these two transformations are non-equivalent as they produce different outcomes and thus illustrate ``transformation contextuality''. Specifically,
\begin{eqnarray}
  (x,y,z) &\substack{\rightarrow\\T_1}& \frac{1}{4}\left[(x,y,z)+(x,-y,-z)\right.\\ && \quad \left.+(-x,y,-z)+(-x,-y,z)\right]. \nonumber
\end{eqnarray}
Thus \(T_1\) maps ontic states with even (odd) sign parity to ontic states with even (odd) sign parity. On the other hand
\begin{eqnarray}
  (x,y,z) &\substack{\rightarrow\\T_2}& \frac{1}{4}\left[(-x,-y,-z)+(-x,y,z)\right.\\ && \quad \left.+(x,-y,z)+(x,y,-z)\right]. \nonumber
\end{eqnarray}
\(T_2\) maps ontic states with even (odd) sign parity to ontic states with odd (even) sign parity. These two sets of four points are different and therefore the two maps can produce different probability distributions, as shown in Fig.~\ref{fig:8stateonticspace}.

As a result, this model produces different probability distributions over the ontic states depending on whether \(T_1\) or \(T_2\) is taken. However, since both maps result in the same result---the maximally mixed state---the resultant probability distribution should be the same. Thus, the eight-state model exhibits transformation contextuality. Lillystone \emph{et al.} then prove that every one-qubit non-preparation contextual ontological model can be mapped to the eight-state model and so all such models exhibit transformation contextuality. This includes both \(\psi\)-epistemic and \(\psi\)-ontic ontological models. We examine their proof carefully in Section~\ref{sec:inequivalence}.

\section{The Blowtorch Map in the Grassmann WWM Formalism}
\label{sec:cptpmapinwwmformalism}

The qubit Wigner-Weyl-Moyal (WWM) formalism was originally introduced by Berezin~\cite{Berezin77} and fully developed in~\cite{Kocia17_2}. The Grassmann model at \(\mathcal O(\hbar^0)\) provides a classical Hamiltonian system that yields spin-\(\frac{1}{2}\) under canonical quantization. It makes use of \(\xi_p\), \(\xi_q\) and \(\xi_r\), three real generators of a Grassmann algebra \(\mathcal G_3\) which obey the anticommutation relation:
\begin{equation}
  \xi_j \xi_k + \xi_k \xi_j \equiv \{\xi_j, \xi_k\} = 0,  \quad \text{for} \, j,k \in \{p,q,r\}.
\end{equation}

Any element \(g \in \mathcal G_3\) may be represented as a finite sum of homogeneous monomials of the Grassmann elements and \(g\) is called a Weyl symbol.

In an effort to examine the \(T_1\) and \(T_2\) maps in this qubit WWM hidden variable theory, we consider the Weyl symbol of a single qubit pure state \(\hat \rho\):
\begin{equation}
  \label{eq:weylsymbolsinglequbit}
  \rho = \frac{1}{2}\left(1 + \alpha i \xi_r \xi_q + \beta i \xi_p \xi_q + \gamma i \xi_p \xi_r\right),
\end{equation}
where \(\alpha^2 + \beta^2 + \gamma^2 = 1\), for \(\alpha, \, \beta, \, \gamma \in \mathbb R\). The \(i\)'s make the Weyl symbol \(\rho\) real, under a generalized conjugation operation~\cite{Kocia17_2}.

Transformations \(\hat I \hat \rho \hat I\), \(\hat X \hat \rho \hat X\), \(\hat Y \hat \rho \hat Y\), \(\hat Z \hat \rho \hat Z\), and \(\hat H \hat \rho \hat H\) are all Clifford transformations and so can be captured in the Wigner-Weyl-Moyal formalism at order \(\hbar^0\) by solving the following classical equations of motion:
\begin{equation}
  \label{eq:eomGrassmann}
  \frac{\mbox d}{\mbox d t} \xi_k = \{H,\xi_k\}_{\text{P.B}} = i H \frac{\cev \partial}{\partial \xi_k}.
\end{equation}
where the right derivative \(\frac{\cev \partial}{\partial \xi_k}\) is as defined in~\cite{Kocia17_2}, \(H_I = 1\), \(H_X = -i \xi_r \xi_q\), \(H_Y = -i \xi_p \xi_q\), \(H_Z = -i \xi_p \xi_r\), for \(t=\pi/2\) and \(H_{\hat H} = -\frac{i}{\sqrt{2}} ( \xi_r \xi_q + \xi_p \xi_r)\) for \(t=\pi\). For \(\hbar > 0\), these equations of motion are deformed to the Weyl algebra~\cite{Kocia17_2} for non-Clifford unitaries. They are then described by a Weyl bracket instead of a Poisson bracket and the Grassmann elements become the usual Pauli matrices in quantum mechanics. However, since this is unnecessary for Clifford transformations, we will not need to explore this regime.

Clifford transformations take stabilizer states to stabilizer states. Solving the equations of motion for transformations \(I\), \(X\), \(Y\), \(Z\), and \(H\), can be written in the same way as the \((x,y,z)\) transformations in~\cite{Lillystone18} by using \(3\)-tuples \((x,y,z) \in \Lambda_2\) for \(x\), \(y\), \(z \in \{\pm \xi_p, \pm \xi_q \pm \xi_r\}\):
\begin{equation}
\label{eq:grassmannIdentitytrans}
  (\xi_p,\xi_q,\xi_r)   \underset{I}{\rightarrow} (\xi_p,\xi_q,\xi_r),
\end{equation}
\begin{equation}
  \label{eq:grassmannPaulitransX}
  (\xi_p,\xi_q,\xi_r) \underset{X}{\rightarrow} (\xi_p,-\xi_q,-\xi_r),
\end{equation}
\begin{equation}
  (\xi_p,\xi_q,\xi_r) \underset{Y}{\rightarrow} (-\xi_p,\xi_q,-\xi_r),
\end{equation}
\begin{equation}
  \label{eq:grassmannPaulitransZ}
  (\xi_p,\xi_q,\xi_r) \underset{Z}{\rightarrow} (-\xi_p,-\xi_q,\xi_r),
\end{equation}
and
\begin{equation}
  \label{eq:grassmannHtrans}
  (\xi_p,\xi_q,\xi_r) \underset{H}{\rightarrow} (\xi_q,\xi_p,-\xi_r).
\end{equation}

Substituting in the maps given by Eqs.~\ref{eq:grassmannIdentitytrans}-\ref{eq:grassmannHtrans}, we find that \(\hat I \hat \rho \hat I\), \(\hat X \hat \rho \hat X\), \(\hat Y \hat \rho \hat Y\), \(\hat Z \hat \rho \hat Z\), and \(\hat H \hat \rho \hat H\) are
\begin{equation}
\label{eq:identitytransformWeyl}
  \rho_I = \frac{1}{2}\left(1 + \alpha i \xi_r \xi_q + \beta i \xi_p \xi_q + \gamma i \xi_p \xi_r\right).
\end{equation}
\begin{equation}
  \rho_X = \frac{1}{2}\left(1 + \alpha i \xi_r \xi_q - \beta i \xi_p \xi_q - \gamma i \xi_p \xi_r\right).
\end{equation}
\begin{equation}
  \rho_Y = \frac{1}{2}\left(1 - \alpha i \xi_r \xi_q + \beta i \xi_p \xi_q - \gamma i \xi_p \xi_r\right),
\end{equation}
\begin{equation}
  \rho_Z = \frac{1}{2}\left(1 - \alpha i \xi_r \xi_q - \beta i \xi_p \xi_q + \gamma i \xi_p \xi_r\right),
\end{equation}
and
\begin{equation}
\label{eq:HadtransformWeyl}
  \rho_{H} = \frac{1}{2}\left(1 + \gamma i \xi_r \xi_q - \beta i \xi_p \xi_q + \alpha i \xi_p \xi_r\right),
\end{equation}
respectively.

Thus, we see that under the \(T_1\) transformation,
\begin{eqnarray}
  \rho &\underset{T_1}{\rightarrow}& \frac{1}{4} \left[ \frac{1}{2}\left(1 + \alpha i \xi_r \xi_q + \beta i \xi_p \xi_q + \gamma i \xi_p \xi_r\right)\right.\\
  && + \frac{1}{2}\left(1 + \alpha i \xi_r \xi_q - \beta i \xi_p \xi_q - \gamma i \xi_p \xi_r\right) \nonumber\\
  && + \frac{1}{2}\left(1 - \alpha i \xi_r \xi_q + \beta i \xi_p \xi_q - \gamma i \xi_p \xi_r\right) \nonumber\\
  && \left. + \frac{1}{2}\left(1 - \alpha i \xi_r \xi_q - \beta i \xi_p \xi_q + \gamma i \xi_p \xi_r\right) \right] \nonumber\\
  &=& \frac{1}{4}(\rho_I+\rho_X+\rho_Y+\rho_Z) \nonumber\\
  &=& \frac{1}{2}. \nonumber
\end{eqnarray}
This is the Weyl symbol for \(\hat I/2\). The simplification of the convex combination above is accomplished by the Weyl algebra of \(\mathcal G_3\). Such a simplification is not possible under \(\{\pm1\}^3\), which lacks such algebraic operations.

On the other hand, acting on this evolution with the Hadamard gate to effect transformation \(T_2\) produces:
\begin{eqnarray}
  \rho &\underset{T_2}{\rightarrow}& \frac{1}{4}\left[ \frac{1}{2}\left(1 + \gamma i \xi_r \xi_q - \beta i \xi_p \xi_q + \alpha i \xi_p \xi_r\right)  \right.\\
       && + \frac{1}{2}\left(1 + \gamma i \xi_r \xi_q + \beta i \xi_p \xi_q - \alpha i \xi_p \xi_r\right) \nonumber\\
       && + \frac{1}{2}\left(1 - \gamma i \xi_r \xi_q - \beta i \xi_p \xi_q - \alpha i \xi_p \xi_r\right) \nonumber\\
       && \left.+ \frac{1}{2}\left(1 - \gamma i \xi_r \xi_q + \beta i \xi_p \xi_q + \alpha i \xi_p \xi_r\right) \right] \nonumber\\
       &=& \frac{1}{4} \left( \rho_H + \rho_{HXH} + \rho_{HYH} + \rho_{HZH} \right) \nonumber\\
       &=& \frac{1}{2}. \nonumber
\end{eqnarray}
Again, this is the Weyl symbol for \(\hat I/2\). Both of these results are obtained without quantizing the Weyl symbols and so this result is possible all while working at order \(\hbar^0\).

This result raises an interesting question when compared to the result obtained using Lillystone \emph{et al}.'s eight-state ontological model: since the WWM formalism is able to obtain the maximally mixed state at order \(\hbar^0\) regardless of whether map \(T_1\) or \(T_2\) is taken, does this suggest that there exists an analogous classical probability theory (a preparation non-contextual ontological model) that similarly does not depend on whether transformation \(T_1\) or \(T_2\) is taken? If so, how can this be reconciled with Lillystone \emph{et al}.'s proof that every such ontological model can be mapped to their eight-state ontological model, which does exhibit dependence on whether \(T_1\) or \(T_2\) is taken?

We investigate these questions in the following sections by first defining a simple three-state ontological model example in Section~\ref{sec:example}, which introduces the key element that the eight-state ontological model does not possess: non-disjoint ontic states. This then leads us to develop a larger ontological model equivalent to the Grassmann WWM formalism in Section~\ref{sec:grassmannHVT} and \ref{sec:fundamentals}.

\section{Example of a Simple Ontological Model with Non-Disjoint Ontic States}
\label{sec:example}

The eight-state model is an example of an ontological model with disjoint ontic states. This means that for any two ontic states \(A\) and \(B\), \(P(A \cup B) = P(A) + P(B)\). By contradistinction, non-disjoint ontic states have non-zero overlaps (\(A \cap B \ne \emptyset\)) and so satisfy the classical relation: \(P(A \cap B) = P(A) + P(B) - P(A \cup B)\). This can be derived directly from Kolmogorov's three axioms as we noted in Section~\ref{sec:review}.

Here we introduce a simple example of a classical probability theory that is defined over only three ontic states, two of which are non-disjoint due to an additional set of relations that satisfies Kolmogorov's axioms. Within this simple model, we show how re-expressing the ontic states in terms of only disjoint states does not produce a convex set of probability distributions due to this additional set of constraints.

Consider a probability space with three elements, \(\Lambda =\{A,\, B,\, C\}\). We wish to deal with a proper probability space, and so must satisfy all of Kolmogorov's axioms given in Section~\ref{sec:review}.

We specify additional constraints on our probability space that we will show are compatible with these axioms:
\begin{equation}
  \label{eq:nondisjointeq1}
  P(C) = 1,
\end{equation}
\begin{equation}
  P(A) + P(B) = 1,
\end{equation}
\begin{equation}
  \label{eq:nondisjointeq3}
  P(A \cup B) = \max\{P(A), P(B)\},
\end{equation}
and
\begin{equation}
  \label{eq:nondisjointeq4}
  P(A \cap B) = \min\{P(A), P(B)\}.
\end{equation}
These additional constraints impose that our ontic states \(A\) and \(B\) are disjoint.

Axiom \(1\) is satisfied since \(\Lambda = C\) and \(P(C) = 1\) and axiom \(2\) can be imposed.

Axiom \(3\) is satisfied since \(A\) and \(B\) are not disjoint and so satisfy the sum rule:
\begin{equation}
  P(A \cup B) = P(A) + P(B) - P(A \cap B).
\end{equation}

\begin{figure}[ht]
  \includegraphics{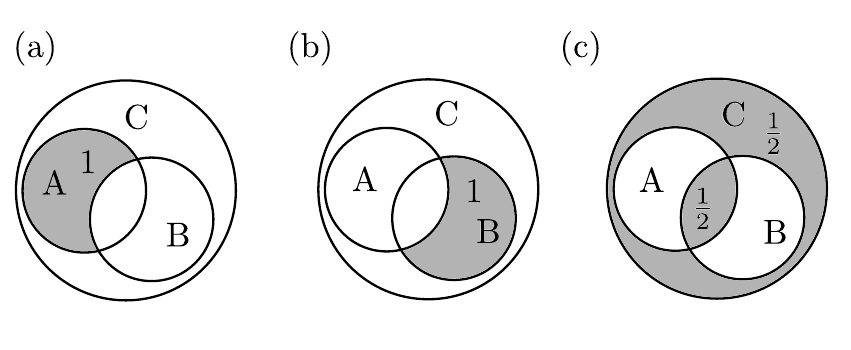}
  \caption{Probability distributions where \(P(C)=1\) and (a) \(P(A)=1\), \(P(B)=0\), (b) \(P(A)=0\), \(P(B)=1\) and (c) \(P(A)=\frac{1}{2}\), \(P(B)=\frac{1}{2}\) that are supported on the the non-disjoint states \(A\) and \(B\) and so satisfy Eqs.~\ref{eq:nondisjointeq1}-\ref{eq:nondisjointeq4}. }
  \label{fig:exampleprobdistros}
\end{figure}

Since \(P(C)=1\), all probability distributions only cover a part of our ontic space. We show three example probability distributions in Fig.~\ref{fig:exampleprobdistros} that satisfy the additional constraints given by Eqs.~\ref{eq:nondisjointeq1}-\ref{eq:nondisjointeq4}.

\begin{figure}[ht]
  \includegraphics{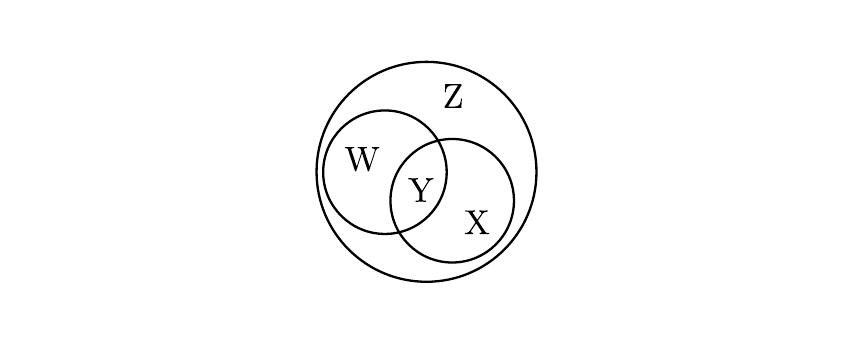}
  \caption{Same ontic space as in Fig.~\ref{fig:onticspace} but now labelled by a ``finer'' set of disjoint ontic states \(W\), \(X\), \(Y\) and \(Z\).}
  \label{fig:disjointonticspace}
\end{figure}

It is of course perfectly acceptable to split up our ontic space into four ``finer'' disjoint ontic states~\cite{Wallman12,Lillystone18,Karanjai18}, which we label \(W\), \(X\), \(Y\) and \(Z\) as in Fig.~\ref{fig:disjointonticspace}.

However, while with the ``coarse-grained'' non-disjoint states (\(A\), \(B\) and \(C\)) the probability distributions form a single convex set, with the ``atomic'' or ``finer'' disjoint states (\(W\), \(X\), \(Y\) and \(Z\)), the additional set of constraints splits this convex set into more than one subset.

To see this, note that for the probability space labelled by the disjoint ontic states \(W\), \(X\), \(Y\), and \(Z\), incorporating the additional system of equations given by Eqs~\ref{eq:nondisjointeq1}-\ref{eq:nondisjointeq4} produces:
\begin{equation}
\label{eq:disjointeq1}
    P(W)+P(X)+P(Y)+P(Z)=1,
\end{equation}
\begin{equation}
\label{eq:disjointeq2}
    P(W)+P(X)+2P(Y)=1,
\end{equation}
\begin{equation}
\label{eq:disjointeq3}
    P(W)+P(X)+P(Y) = \max\{P(W)+P(Y),P(X)+P(Y)\},
\end{equation}
\begin{equation}
\label{eq:disjointeq4}
    P(Y) = \min\{P(W)+P(Y),P(X)+P(Y)\},
\end{equation}
respectively.

Allowed probability distributions are points in the three-simplex defined by Eq.~\ref{eq:disjointeq3} that also satisfy Eqs~\ref{eq:disjointeq1},~\ref{eq:disjointeq2}, and~\ref{eq:disjointeq4}. There are only two cases of solutions:
\begin{enumerate}
    \item \(P(W)=0\),
    \item \(P(X)=0\).
\end{enumerate}

Let the tuple \((w,x,y,z) \in \Lambda_3\) refer to the probability on \(W\), \(X\), \(Y\), and \(Z\), respectively. In cases \(1\) and \(2\) we can choose \(\alpha \equiv P(W)\) or \(\alpha \equiv P(X)\) respectively, and define a one-parameter family of probability distributions \((w,x,y,z)\):
\begin{flalign}
    \label{eq:threeonticstatesconvexset1} \indent 1. \quad (\alpha,0,\frac{1}{2}(1-\alpha),\frac{1}{2}(1-\alpha)), &&\\
    \label{eq:threeonticstatesconvexset2} \indent 2. \quad (0,\alpha,\frac{1}{2}(1-\alpha),\frac{1}{2}(1-\alpha)), &&
\end{flalign}
respectively, for \(0 \le \alpha \le 1\). \(\alpha = 0\) corresponds to the only probability distribution that lies in both cases and is the one indicated by Fig.~\ref{fig:exampleprobdistros}c. These two cases correspond to two convex subsets of the original single convex set.

Convex combination of probability distributions from these two convex sets do \emph{not} satisfy the constraints given by Eqs.~\ref{eq:disjointeq1}-\ref{eq:disjointeq4} (unless \(\alpha = 0\)). For instance, consider the convex combination of the probability distributions in Fig.~\ref{fig:exampleprobdistros}a and b corresponding to the following tuples in \(\Lambda_3\):
\begin{equation}
\label{eq:exampleconvexcombo}
  \frac{1}{2}\bigg(1,0,0,0\bigg)+\frac{1}{2}\bigg(0,1,0,0\bigg).
\end{equation}

For the probability space labelled by disjoint ontic states \(W\), \(X\), \(Y\), and \(Z\), since the two terms in the convex combination given by Eq.~\ref{eq:exampleconvexcombo} correspond to two different cases (\(1\) and \(2\)) with \(\alpha \ne 0\), it follows that their result cannot satisfy Eqs.~\ref{eq:disjointeq1}-\ref{eq:disjointeq4}. The only way to obtain a convex combination is to convert the disjoint ontic states \(W\), \(X\), \(Y\), and \(Z\), back into the non-disjoint ontic states \(A\), \(B\) and \(C\), perform the convex combination that satisfies the old Eqs.~\ref{eq:nondisjointeq1}-\ref{eq:nondisjointeq4}, and then convert back to the disjoint ontic states.

Doing so, we can find that the convex combination given by Eq.~\ref{eq:exampleconvexcombo}, when converted to be in terms of non-disjoint states \(A\) and \(B\), produces \(P(A=W \cup Y)=P(B = X\cup Y)=\frac{1}{2}\). From Eqs~\ref{eq:nondisjointeq3}-\ref{eq:nondisjointeq4}, this means that \(P(A \cap B) = P(A \cup B) = \frac{1}{2}\). Moreover, by Eq.~\ref{eq:nondisjointeq1}, the resultant probability distribution must have zero support on \(A \setminus B\) and \(B \setminus A\). Converting back to the disjoint states \(W\), \(X\), \(Y\), \(Z\), this means that \(P(W=A \setminus B) = P(X=B \setminus A) = 0\) and \(P(Y=A \cap B) = P(Z=A \cup B) = \frac{1}{2}\). This is represented by the tuple \((0,0,\frac{1}{2},\frac{1}{2}) \in \Lambda_3\), which is the probability distribution in Fig.~\ref{fig:exampleprobdistros}c.

In other words, given the information that there is a probability \(\frac{1}{2}\) of being found in \(A\) and a probability \(\frac{1}{2}\) of being found in \(B\), this model enforces that \(A\) and \(B\) are non-disjoint and so produces the physically intuitive result that the probability of being found in \(A\) or \(B\) is \(\frac{1}{2}\) (and so the probability of being found in neither \(A\) or \(B\) is \(\frac{1}{2}\) too). This is a very different outcome from the one obtained if \(A\) and \(B\) are assumed to be disjoint, which given the information that there is a probability \(\frac{1}{2}\) of being found in \(A\) and a probability \(\frac{1}{2}\) instead implies that the probability of being found in \(A\) or \(B\) is \(1\).

Though the constraints given by Eqs.~\ref{eq:nondisjointeq1}-\ref{eq:nondisjointeq4} produce a single convex set with the ``coarse'' set of non-disjoint states \(A\), \(B\), and \(C\), the ``finer'' disjoint set \(W\), \(X\), \(Y\), and \(Z\), cannot satisfy them with a single convex set. 

Indeed, additional relations can only \emph{non-trivially} supplement Kolmogorov's axioms if they produce two or more convex subsets when the ontic states are expressed disjointly (with no overlaps). This is because additional relations that satisfy \(\sigma\)-additivity for all ontic states (i.e. all ontic states are disjoint) add nothing new to the probability theory \emph{unless} they produce more than one convex subset. However, for the theory to still describe the subtheory of interest, i.e. for the additional relations not to be too constraining, there must exist some other set of (non-disjoint) ontic states with respect to which all the probability distributions fall into the same convex set. This example demonstrates that such a middle ground between ``unconstrained'' ontological models, which produce one convex set regardless of which set of disjoint or non-disjoint ontic states they are expressed with, and ``overconstrained'' ontological models, which produce more than one convex set regardless of which set of ontic states they are expressed with, exists. This middle ground consists of constrained ontological models, which produce one convex set with respect to a particular set of non-disjoint ontic states and more than one for all other sets. This possibility appears to have been overlooked in the literature.

\section{Grassmann WWM as an Ontological Model}
\label{sec:grassmannHVT}

In our prior work~\cite{Kocia17_2} we showed that it is possible to construct a local hidden variable theory (an ontological model) from the Grassmann WWM formalism to describe qubit stabilizer propagation using a non-negative probability distribution defined over states corresponding to the Grassmann monomials \(\xi_j \xi_k\). We now re-present these results with respect to the nomenclature used to examine the simple ontological model in Section~\ref{sec:example}.

A measure on the \(\mathcal G_3\) algebra can be defined for any state \(\rho=\ketbra{\psi}{\psi}\),
\begin{equation}
  \mu_{\rho}(A_i) = \int \rho(\xi) \tilde A(\xi) \text d^3 \bs \xi,
\end{equation}
where \(\tilde A(\xi)\) is the dual (odd) Weyl symbol of \(A(\xi)\)~\cite{Kocia17_2}. When \(A(\xi) \equiv A_i(\xi)\) is the Weyl symbol of an element of a positive-operator valued measure (POVM) \(\hat A_i\), then \(\mu_\rho(A_i)\) is a non-negative measure (a probability distribution) over outcomes \(A_i\) of state \(\rho\):
\begin{equation}
    P_{A_i}(\rho) \equiv \mu_{\rho}(A_i).
\end{equation}

However, we cannot rely on the measure \(\mu_{\rho}(A)\) as a probability measure over ontic states \(\xi_i \xi_j\) since it can be negative for \(A\in\{\xi_j \xi_k\}\) as they are not elements of POVMs. Nevertheless, we can define a one-to-one map between \(\mu_{\rho}(\xi_j \xi_k)\) and a \emph{bone fide} probability measure that also preserves convex combination if we consider the Weyl algebra that the \(\xi_j \xi_k\) satisfy.

For the ontic state \(\xi_j \xi_k\),
\begin{widetext}
\begin{equation}
  -\frac{1}{2} \le \mu_\rho(\xi_j \xi_k) = \int \xi_j \xi_k \tilde \rho(\xi) \text d^3 \bs \xi = \frac{1}{2} \sum_l \epsilon_{jkl} (\alpha_l + \beta_l + \gamma_l) \le \frac{1}{2},
\end{equation}
\end{widetext}
for all one-qubit states \(\rho\).

We note that
\begin{equation}
    \mu_\rho(\xi_j \xi_k) = \mu_\rho(-\xi_k \xi_j) = - \mu_\rho(\xi_k \xi_j),
\end{equation}
and so \(\mu_\rho(\xi_k \xi_j) < 0\) is the same statement as \(\mu_\rho(\xi_j \xi_k) > 0\). We choose to interpret \(\mu_\rho(\xi_j \xi_k) > 0\) as proportional to the non-negative measure of ontic state \(\xi_j \xi_i \setminus \xi_i \xi_j\) (and vice-versa). 

Given a probability \(P(\xi_j \xi_k \setminus \xi_k \xi_j) \propto \mu_\rho(\xi_j \xi_k) > 0\), we further choose the probability of the other ontic state, \(P(\xi_k \xi_j \setminus \xi_j \xi_k)\), to be zero under the heuristic motivation that \(\mu_\rho\) does not need to track something if it is zero. Thus, given an ontic state \(\lambda \in \{\xi_j \xi_k\}\), we define the non-negative probability of stabilizer state \(\rho\) in ontic state \(\lambda \setminus - \lambda\) to be
\begin{equation}
\label{eq:proboflambdainrho}
  P_\rho(\lambda \setminus -\lambda) \equiv \max\{ 2\mu_\rho(\lambda), 0\},
\end{equation}
where the factor of \(2\) allows the probability to saturate an upper bound of \(1\). We note that this is perhaps an arbitrary definition, we shall see that it is an acceptable one as it produces a theory consistent with Kolmogorov's axioms once unions and intersections are included, and reproduces the Grassmann WWM formalism for the stabilizer subtheory.

Any single qubit state's Weyl symbol \(\rho\) is represented by a linear combination of Grassmann monomials as in Eq.~\ref{eq:weylsymbolsinglequbit}. Thus, our choice of definition for \(P_\rho\) equates the ``addition'' operator in the Weyl algebra to a ``convex addition'' operator since it treats any linear combination involving negative coefficients in front of Grassmann monomials as a unique non-negative convex combination, making use of the Grassmann anticommutation relations.

A stabilizer state has the Weyl symbol
\begin{equation}
\label{eq:stabstateWeylsymbol}
    \rho_{jk} \equiv \frac{1}{2}(1 + i \xi_j \xi_k).
\end{equation}
We now consider a convex combination of the two distinct stabilizer states \(\rho_{jk}\) and \(\rho_{kj}\) under the Weyl algebra:
\begin{eqnarray}
\rho &=& \alpha \rho_{jk} + \beta \rho_{kj} \\
    &=& \frac{1}{2} + (\alpha - \beta) i \xi_j \xi_k, \nonumber
\end{eqnarray}
for \(\alpha\), \(\beta \ge 0\) such that \(\alpha + \beta = 1\). Note that \((\alpha - \beta) i \xi_j \xi_k = (\beta - \alpha) i \xi_k \xi_j\).

WLOG, let us assume that \(\alpha \ge \beta\). 
Eq.~\ref{eq:proboflambdainrho} for \(P_\rho(\lambda \setminus -\lambda)\) means that the probability of being in ontic state \(\xi_j \xi_k \setminus \xi_k \xi_j\) after this convex combination is two times the coefficient in front of the resultant Weyl symbol's \(\xi_j \xi_k\) term, \(\alpha - \beta\), and the probability of being in ontic state \(\xi_k \xi_j \setminus \xi_j \xi_k\) is \(0\). Before we simplified the convex combination, \(P(\xi_j \xi_k) = \alpha\) and so 
\begin{equation}
    P_\rho(\xi_j \xi_k \cap \xi_k \xi_j) = P_\rho(\xi_j \xi_k) - P_\rho(\xi_j \xi_k \setminus \xi_k \xi_j) = \beta = \min\{\alpha,\beta\}.
\end{equation}
This further agrees with
\begin{equation}
    P_\rho(\xi_j \xi_k \cap \xi_k \xi_j) = P_\rho(\xi_k \xi_j) - P_\rho(\xi_k \xi_j \setminus \xi_j \xi_k) = \beta = \min\{\alpha,\beta\},
\end{equation}
since \(P_\rho(\xi_k \xi_j) = \beta\). In other words, the convex combination takes a probability density of \(\min\{\alpha,\beta\}\) from ontic state \(\xi_j \xi_k \setminus \xi_k \xi_j\) to the intersection between the two ontic states. This means that
\begin{equation}
    P(\xi_j \xi_k \cup \xi_k \xi_j) = P_\rho(\xi_j \xi_k \setminus \xi_k \xi_j) + P_\rho(\xi_k \xi_j \setminus \xi_j \xi_k) = \alpha = \max\{\alpha,\beta\}.
\end{equation}

Therefore, for a map between \(\mu\) and the probabilities to preserve \(\mu\)'s convex combinations under its Weyl algebra, it follows that
\begin{equation}
    P(\xi_j \xi_k \cup \xi_k \xi_j) = \max\{P(\xi_j \xi_k), P(\xi_k \xi_j)\}
\end{equation}
and
\begin{equation}
    P(\xi_j \xi_k \cap \xi_k \xi_j) = \min\{P(\xi_j \xi_k), P(\xi_k \xi_j)\}.
\end{equation}

As a result, we have the same probability space as that considered in the simple example of Section~\ref{sec:example}, except that instead of one independent pair \(A\) and \(B\), we have three independent pairs. Moreover, we accomplished this via a one-to-one mapping between our probabilities and our measure \(\mu\) in \(\mathcal G_3\) such that the set of probability distributions, when considered over the non-disjoint \(\xi_j \xi_k\) ontic states, is a convex set. Most importantly, as we showed in the previous section, these additional constraints satisfy Kolmogorov's axioms and so form a valid classical probability theory or ontological model.

Using non-disjoint ontic states, we can set \(A_1 = \xi_{pq}\), \(B_1 = \xi_{qp}\), and then add two additional pairs: \(\{A_2 = \xi_{pr}, B_2 = \xi_{rp}\}\) and \(\{A_3 = \xi_{qr}, B_3 = \xi_{rq}\}\) so that:
\begin{equation}
\label{eq:nondisjointthreepairseq1}
  P(A_i) + P(B_i) = 1,
\end{equation}
\begin{equation}
  P(A_i \cup B_i) = \max\{P(A_i), P(B_i)\},
\end{equation}
and
\begin{equation}
  \label{eq:nondisjointthreepairseq3}
  P(A_i \cap B_i) = \min\{P(A_i), P(B_i)\}.
\end{equation}
\(C = \Lambda\) now, and \(P(\Lambda)=1\) is enforced by Kolmogorov's first axiom.

The allowed probability distributions all belong in the same family and for \((a_1, a_2, a_3, b_1, b_2, b_3) \in \Lambda_4\), where \(a_1\) is the probability to be in \(A_1\) and so on, they take the form:
\begin{equation}
    (\alpha,\beta,\gamma,1-\alpha,1-\beta,1-\gamma),
\end{equation}
where \(0 \le \alpha, \beta, \gamma \le 1\).

Since there are no relations that govern the probabilities between the different pairs, these three sets of ontic states (\(1\), \(2\), and \(3\)) are independent of each other. Convex combinations of any probability distribution defined on these disjoint ontic states produce another probability distribution on the disjoint ontic states that satisfies the constraints given by Eqs.~\ref{eq:nondisjointthreepairseq1}-\ref{eq:nondisjointthreepairseq3}; there is only one convex set of probability distributions.

On the other hand, using disjoint ontic states, we can set \(W_1 = \xi_{rq}\setminus\xi_{qr}\), \(X_1 = \xi_{qr}\setminus\xi_{rq}\), \(Y_1 = \xi_{rq} \cap \xi_{qr}\), and \(Z_1 = (\xi_{rq} \cup \xi_{qr})^c\), and then add two additional pairs: \(\{W_2 = \xi_{pq}\setminus\xi_{qp}, X_2 = \xi_{qp}\setminus\xi_{pq}\}\) and \(\{W_3 = \xi_{pr}\setminus\xi_{rp}, X_3 = \xi_{rp}\setminus\xi_{pr}\}\), where we define \(Y_3\), \(Z_3\), \(Y_4\), and \(Z_4\) in a similar manner.

\(W_i\), \(X_i\), \(Y_i\), and \(Z_i\) satisfy all the constraints that \(W\), \(X\), \(Y\), and \(Z\) did:
\begin{equation}
\label{eq:disjointthreepairseq1}
  P(W_i) + P(X_i) + P(Y_i) = \max \{P(W_i) + P(Y_i), P(X_i) + P(Y_i)\},
\end{equation}
\begin{equation}
    P(Y_i) = \min \{P(W_i) + P(Y_i), P(X_i) + P(Y_i)\},
\end{equation}
\begin{equation}
  P(W_i) + P(X_i) + P(Y_i) + P(Z_i) = 1,
\end{equation}
\begin{equation}
\label{eq:disjointthreepairseq4}
  P(W_i) + P(X_i) + 2 P(Y_i) = 1.
\end{equation}

Now there are \(2^3 = 8\) families of solutions that satisfy Eqs.~\ref{eq:disjointthreepairseq1}-\ref{eq:disjointthreepairseq4}. As before in Eqs.~\ref{eq:threeonticstatesconvexset1}-\ref{eq:threeonticstatesconvexset2}, we can find that \(P(Z_i) = P(Y_i)\) and so we discard \(P(Z_i)\) when listing these \(8\) cases \((w_1,w_2,w_3,x_1,x_2,x_3,y_1,y_2,y_3) \in \Lambda_5\), where \(w_1\) is the probability of being in \(W_1\) and so on. The set of solutions corresponds to all possible permutations of the two solutions given in Eqs.~\ref{eq:threeonticstatesconvexset1}-\ref{eq:threeonticstatesconvexset2} extended to three independent pairs:
\begin{flalign}
\label{eq:threepairdisjointsolns1}
  \indent 1. \quad (\alpha,\beta,\gamma,0,0,0,\frac{1}{2}(1-\alpha),\frac{1}{2}(1-\beta),\frac{1}{2}(1-\gamma)), &&\\
  \indent 2. \quad (\alpha,\beta,0,0,0,\gamma,\frac{1}{2}(1-\alpha),\frac{1}{2}(1-\beta),\frac{1}{2}(1-\gamma)), &&\\
  \indent 3. \quad (\alpha,0,\gamma,0,\beta,0,\frac{1}{2}(1-\alpha),\frac{1}{2}(1-\beta),\frac{1}{2}(1-\gamma)), &&\\
  \indent 4. \quad (\alpha,0,0,0,\beta,\gamma,\frac{1}{2}(1-\alpha),\frac{1}{2}(1-\beta),\frac{1}{2}(1-\gamma)), &&\\
  \indent 5. \quad (0,\beta,\gamma,\alpha,0,0,\frac{1}{2}(1-\alpha),\frac{1}{2}(1-\beta),\frac{1}{2}(1-\gamma)), &&\\
  \indent 6. \quad (0,\beta,0,\alpha,0,\gamma,\frac{1}{2}(1-\alpha),\frac{1}{2}(1-\beta),\frac{1}{2}(1-\gamma)), &&\\
  \indent 7. \quad (0,0,\gamma,\alpha,\beta,0,\frac{1}{2}(1-\alpha),\frac{1}{2}(1-\beta),\frac{1}{2}(1-\gamma)), &&\\
  \label{eq:threepairdisjointsolns8}
  \indent 8. \quad (0,0,0,\alpha,\beta,\gamma,\frac{1}{2}(1-\alpha),\frac{1}{2}(1-\beta),\frac{1}{2}(1-\gamma)), &&
\end{flalign}
where \(0 \le \alpha, \beta, \gamma \le 1\). These cases only contain one common probability distribution: the distribution \((0,0,\frac{1}{2},0,0,\frac{1}{2},0,0,\frac{1}{2}) \in \Lambda_5\) when \(\alpha = \beta = \gamma = 0\).

Again, these are eight convex subsets of the \(8\)-simplex of all distributions over \(8\) ontic states; convex combinations of the probability distributions above do not satisfy Eqs.~\ref{eq:disjointthreepairseq1}-\ref{eq:disjointthreepairseq4} (unless \(\alpha = \beta = \gamma = 0\)). 

We have thus established that the Grassmann WWM formalism is equivalent to an ontological model defined by three pairs of non-disjoint ontic states for the stabilizer subtheory and produces eight convex subsets when expressed in terms of disjoint ontic states. In the subsequent Section~\ref{sec:fundamentals}, we develop more of its properties.

\section{Properties of the Grassmann WWM Ontological Model}
\label{sec:fundamentals}

In the eight-state model, the ontic space is partitioned into eight disjoint states that are indexed by the eight \(3\)-tuples in \(\Lambda_1\):
\begin{eqnarray}
    \Lambda &=& \{(+,+,+),(+,+,-),(+,-,+),(-,+,+),\\
    &&(+,-,-),(-,+,-),(-,-,+),(-,-,-)\}. \nonumber
\end{eqnarray}
Convex combinations of these eight tuples defines any valid probability distribution in the eight-state model. 

These \(3\)-tuples can be converted into equivalent \(6\)-tuples by defining the \(6\)-tuples to be \((x_+, y_+, z_+, x_-, y_-, z_-) \in \Lambda_6\), where \(x_+ = 1\) and \(x_-=0\) if the first entry of the corresponding \(3\)-tuple is `\(+\)' and \(x_+ = 0\) and \(x_- = 1\) if it the first entry is `\(-\)' and so on. This produces a partition of the ontic space into eight \(6\)-tuples
\begin{eqnarray}
  \label{eq:Grassmannonticspace}
  \Lambda' &=& \{(1,1,1,0,0,0), (1,1,0,0,0,1), \\ && (1,0,1,0,1,0), (1,0,0,0,1,1), (0,1,1,1,0,0), \nonumber\\
  &&(0,1,0,1,0,1), (0,0,1,1,1,0), (0,0,0,1,1,1) \}. \nonumber
\end{eqnarray}
Using \(6\)-tuples (\(\Lambda_6\)) instead of \(3\)-tuples (\(\Lambda_1\)) simplifies the resultant probability distribution of convex combinations because they can now be represented by a single \(6\)-tuple. For instance, the probability distribution \(\frac{1}{2}(+,+,+)+\frac{1}{2}(+,-,-) \in \Lambda_1\) cannot be simplified any further but \(\frac{1}{2}(1,1,1,0,0,0)+\frac{1}{2}(1,0,0,0,1,1) = (1,\frac{1}{2},\frac{1}{2},0,\frac{1}{2},\frac{1}{2}) \in \Lambda_6\).
For general probability distributions, the equation for component-wise convex addition is
\begin{widetext}
\begin{equation}
\label{eq:6tuplecomponentwiseaddition}
    \alpha (x_+, y_+, z_+, x_-, y_-, z_-) + \beta (x'_+, y'_+, z'_+, x'_-, y'_-, z'_-) = (\alpha + \beta) (x_+ + x'_+, y_+ + y'_+, z_+ + z'_+, x_- + x'_-, y_- + y'_-, z_- + z'_-). 
\end{equation}
\end{widetext}

The \(6\)-tuple notation is still useful in simplifying convex combinations of ontic states into a single tuple when applied to the Grassmann WWM ontological model's probability distributions, defined to be \((w_1, w_2, w_3, x_1, x_2, x_3) \in \Lambda_6\). However, now convex combinations of probability distributions must additionally satisfy Eqs.~\ref{eq:nondisjointthreepairseq1}-\ref{eq:nondisjointthreepairseq3} and so the same simple component-wise addition rule of Eq.~\ref{eq:6tuplecomponentwiseaddition} does not hold. 

Nevertheless, the \(6\)-tuple is useful in another way for the Grassmann WWM ontological model because for probability distributions that correspond to quantum states \(\hat \rho\), its entries correspond to the coefficients in front of the ontic states in the Weyl symbol of the state when it is written with the minimal number of terms such that all coefficients are non-negative (a unique form)~\cite{Kocia17_2}:
\begin{equation}
  \label{eq:probdistrogbar}
  \bar g = \left( g_p,\, g_r,\, g_q,\, g_{\text{-}p},\, g_{\text{-}r},\, g_{\text{-}q}\right) \in \Lambda_6,
\end{equation}
where \(g_p = P_\rho(\xi_r\xi_q \setminus \xi_q \xi_r) = \max\{0,2 \mu_g(\xi_r \xi_q)\}\), \(g_{-p} = P_\rho(\xi_q\xi_r \setminus \xi_r \xi_q) = \max\{0,2 \mu_g(\xi_q \xi_r)\}\), etc. Since a stabilizer state \(\rho_\text{stab}\) is given by Eq.~\ref{eq:stabstateWeylsymbol}, and the entries in a \(6\)-tuple in \(\Lambda_6\) correspond to \(P_\rho(\xi_j \xi_k \setminus \xi_k \xi_j)\), the six stabilizer states correspond to the probability distributions,
\begin{eqnarray}
  \label{eq:stabstates}
  &&\rho_\text{stab} \nonumber\\
  &\in& \{(1,0,0,0,0,0), (0,1,0,0,0,0), (0,0,1,0,0,0) \\
  &&(0,0,0,1,0,0), (0,0,0,0,1,0), (0,0,0,0,0,1) \}. \nonumber
\end{eqnarray} Therefore, for stabilizer states the entries in the \(6\)-tuple are five \(0\)s and a single \(1\). This leads to a generalized discrete notion of conserved area or symplecticity for Clifford gates on stabilizer states~\cite{Kocia17_2}. For all these reasons, we will proceed to use this \(6\)-tuple notation from this point onwards.

Note that \((w_1, w_2, w_3, x_1, x_2, x_3) \in \Lambda_6\) uniquely identifies any probability distribution in the Grassmann WWM ontological model since \(y_i\) and \(z_i\) can be determined from \(w_i\) and \(x_i\) (\(y_i = z_i = \frac{1}{2}(1-\max\{w_i,x_i\})\)) as we showed in the last Section and Eqs.~\ref{eq:threepairdisjointsolns1}-\ref{eq:threepairdisjointsolns8}.

The eight ontic states of the eight-state model given by Eq.~\ref{eq:Grassmannonticspace} in the \(6\)-tuple notation, also serve as a valid basis for the convex combination (vector space) operation in the Grassmann WWM ontological model with the \(6\)-tuple appropriately redefined to be \((w_1,w_2,w_3,x_1,x_2,x_3)\). This can be shown by noting that the three sets of ontic states \(W_i\), \(X_i\), \(Y_i\) and \(Z_i\) (or \(A_i\) and \(B_i\)) are independent and that, for \((w_i,x_i)\), convex combinations of \((1,0)\) and \((0,1)\) determine all the possible probability distributions given by Eqs.~\ref{eq:threeonticstatesconvexset1}-\ref{eq:threeonticstatesconvexset2} (after they are converting back to their non-disjoint counterparts, convex added according to the constraints given by Eqs.~\ref{eq:nondisjointeq1}-\ref{eq:nondisjointeq4}, and the converted back to the disjoint \(w_i\) and \(x_i\)). Thus, convex combinations of the Cartesian product \(\{(1,0),(0,1)\}^3\) must determine all the possible probability distributions given by the larger set of three pairs of independent ontic states. This Cartesian product corresponds to the eight states given by the \(6\)-tuples in Eq.~\ref{eq:Grassmannonticspace}.

To find the overlap between a probability distribution \(\rho = (w^\rho_1, w^\rho_2, w^\rho_3, x^\rho_1, x^\rho_2, x^\rho_3)\) and one of the eight ontic states \(\lambda \equiv (w^\lambda_1,w^\lambda_2,w^\lambda_3,x^\lambda_1,x^\lambda_2,x^\lambda_3) \in \Lambda'\) of Eq.~\ref{eq:Grassmannonticspace}, one must be careful to include their probability densities in the intersections \(y_i\) and complements \(z_i\), which as we pointed out, are uniquely determined by \(w_i\) and \(x_i\). In particular, the eight ontic states \(\lambda\) in Eq.~\ref{eq:Grassmannonticspace} have support of \(1\) on three \(w_i\) and/or \(x_i\)s, and \(0\) on all the others. Hence they must have support of \(0\) on all \(y_i\)s and \(z_i\)s (since \(y_i = z_i = \frac{1}{2}(1-\max\{w_i,x_i\})\)).

We have shown that every stabilizer state probability distribution has support of \(1\) on one \(w_i\) or \(x_i\) and \(0\) on all the other \(w_i\)s and \(x_i\)s. This means that each stabilizer state distribution has support of \(\frac{1}{2}\) on two pairs of \(y_i\) and \(z_i\). Therefore, stabilizer state probability distributions have non-zero overlap with the four ontic states in Eq.~\ref{eq:Grassmannonticspace} that also have a `\(1\)' in the same entry of their \(6\)-tuple. For instance,
\begin{eqnarray}
\label{eq:stabstateGrassonticex}
    (1,0,0,0,0,0) &=& \frac{1}{4}\left[(1,1,1,0,0,0)+(1,1,0,0,0,1)\right. \\
    && \left.+(1,0,1,0,1,0)+(1,0,0,0,1,1)\right]. \nonumber
\end{eqnarray}
These actually (superficially) correspond to the same convex combinations as in the eight-state model---as we have seen, the reasoning in the Grassmann WWM ontological involves tallying up additional \(y_i\) and \(z_i\) regions that do not exist in the eight-state model.

According to Definition~\ref{def:epistemic1}, the fact that stabilizer states have support on more than one ontic state means that the Grassmann model is a \(\psi\)-epistemic ontological model for the stabilizer state subtheory; the probability distributions of different (but non-orthogonal) stabilizer states overlap.

\begin{figure}[ht]
  \includegraphics{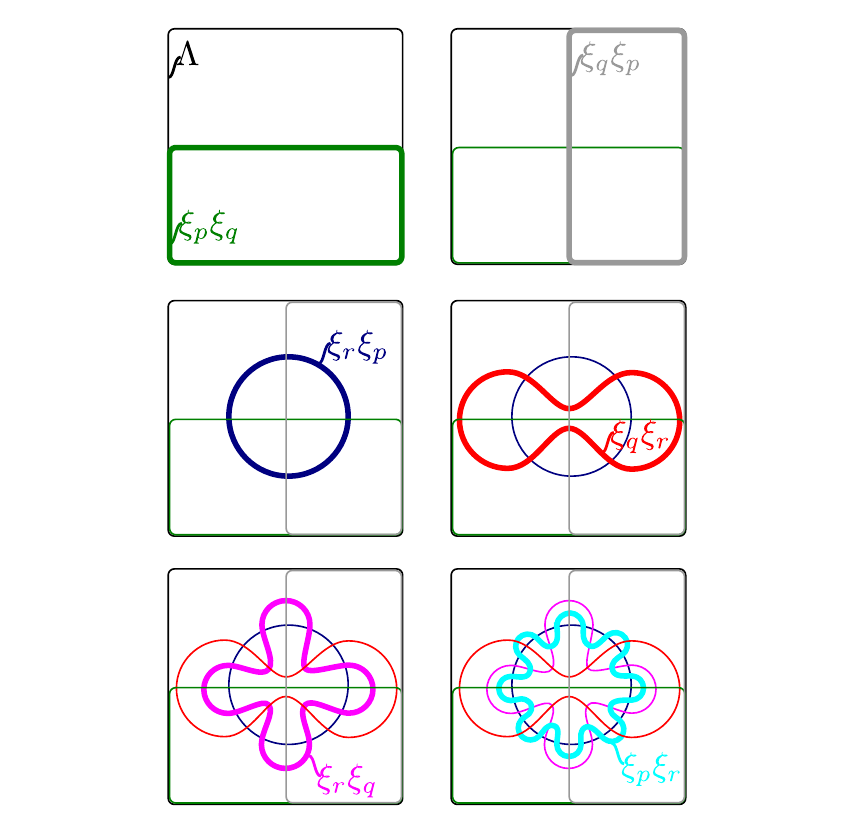}
  \caption{An Edwards-Venn diagram of the ontic space of the Grassmann WWM model that is able to illustrate all the possible overlaps between the non-disjoint ontic states.
  }
  \label{fig:onticspaceVenn}
\end{figure}
 
In summary, even with disjoint ontic states that must be converted to non-disjoint states when taking convex combinations to account for the constraints given by Eqs.~\ref{eq:nondisjointthreepairseq1}-\ref{eq:nondisjointthreepairseq3}, component-wise addition of support on the eight ontic states \(\lambda \in \Lambda'\) given by Eq.~\ref{eq:Grassmannonticspace}, which are equivalent to the eight-state model's, still holds as a way to determine probability distribution overlap.

The Grassmann WWM ontological model is preparation non-contextual for the stabilizer subtheory. Any convex combination of stabilizer state probability distributions produces a unique probability distribution. If two probability distributions are not the same, they do not correspond to the same state since they directly one-to-one map to the Weyl symbol of the state. 

The model is also measurement non-contextual for the stabilizer subtheory since the stabilizer state probability functions given by Eq.~\ref{eq:stabstates} correspond to the conditional probability functions \(\xi_k^M : \Lambda' \rightarrow [0,1]\) of Pauli measurement \(M\), where the probability of outcome \(k\) given measurement \(M\),
\begin{equation}
    Pr(k|M) = \sum_{\lambda \in \Lambda'} \xi_k^M(\lambda) \rho(\lambda).
\end{equation}
This is only the same as the probability of outcome \(k\) under another measurement \(M'\) for all stabilizer states \(\rho\) if the two are equivalent measurements (\(\xi_k^M = \xi_k^{M'} \Leftrightarrow (k,M) \cong (k,M')\)) because no two stabilizer states probability distributions produce the same overlaps with all other stabilizer state probability distributions.

We also demonstrated in our prior work that the Grassmann WWM formalism, and therefore the Grassmann ontological model, exhibits measurement and preparation non-contextuality for one qubit~\cite{Kocia17_2}.

\begin{figure}[ht]
  \includegraphics{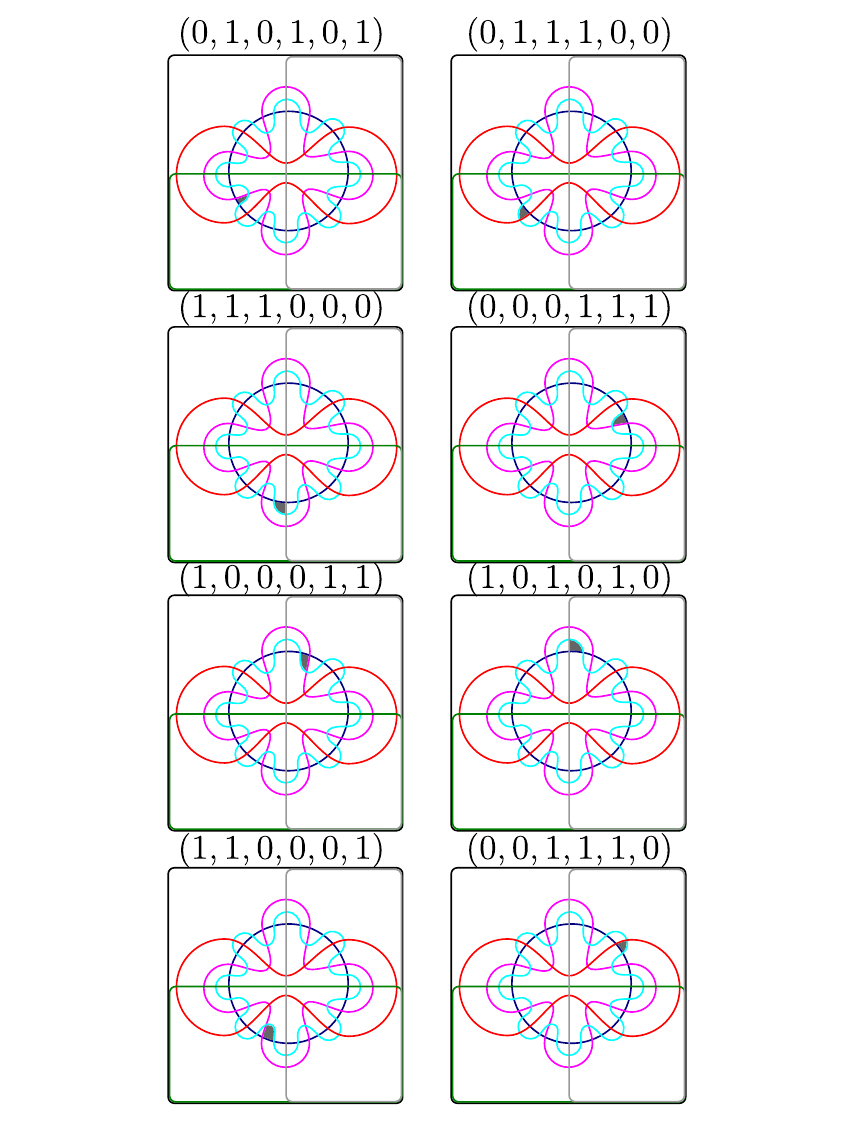}
  \caption{The eight ontic states \(\lambda \in \Lambda'\) of the Grassmann WWM ontological model given by Eq.~\ref{eq:Grassmannonticspace}. Though these disjoint ontic states do not appear to cover all of ontic space, since they satisfy the additional equations given by Eq.~\ref{eq:nondisjointthreepairseq1}-\ref{eq:nondisjointthreepairseq3} when expressed as non-disjoint states, knowledge of the support of a probability distribution on these eight states is sufficient to determine it everywhere else in ontic space.
  }
  \label{fig:onticspace}
\end{figure}

There are a few different ways of illustrating the unions and intersections of the eight ontic states \(\lambda\) given by Eq.~\ref{eq:Grassmannonticspace} in a Venn diagram. We choose to use the Edwards-Venn diagram approach, which takes a hemispheric approach to illustrating overlapping regions between ontic states~\cite{Edwards04}; every additional ontic state added to the Venn diagram has more ``leafs'' or hemispheres that overlap with all previous ontic states thereby capturing all possible combinations of intersections with them. We show our ontic space in the Edwards-Venn diagram of Figure.~\ref{fig:onticspaceVenn} and~\ref{fig:onticspace}.

\begin{figure}[ht]
  \includegraphics{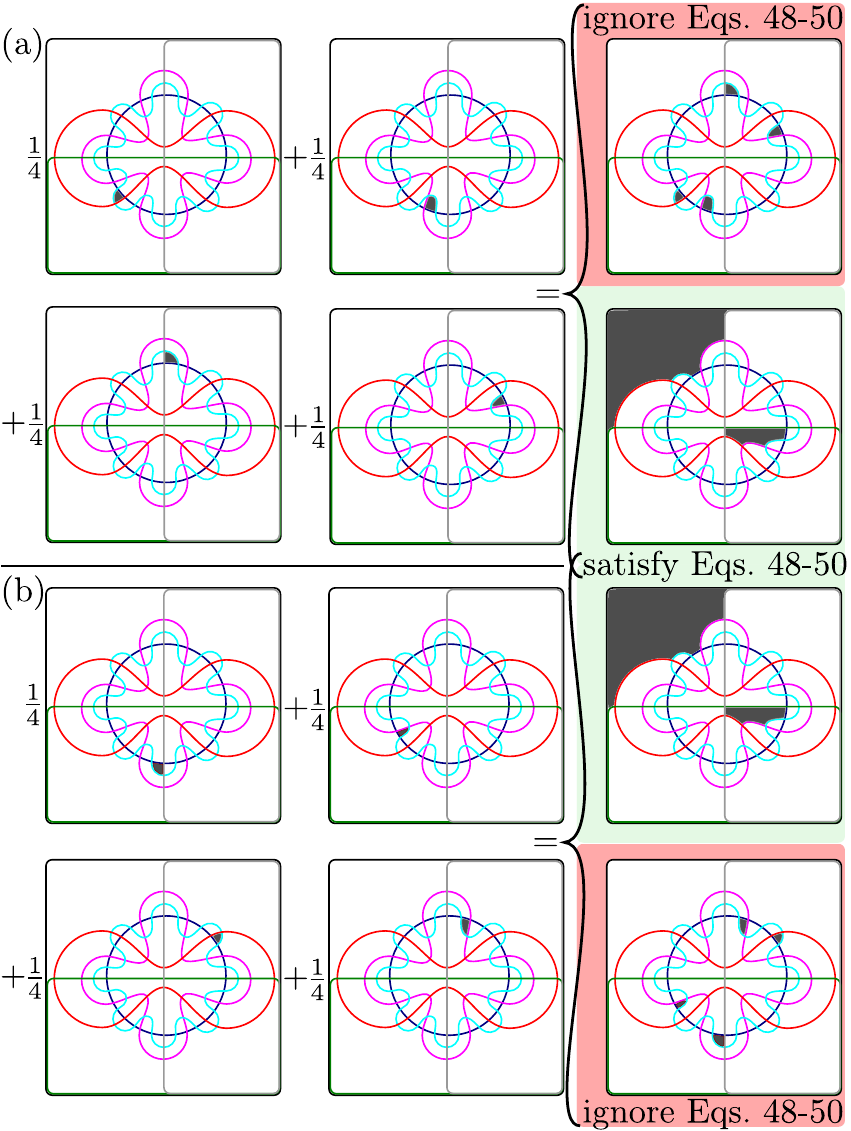}
  \caption{The convex combinations produced by (a) \(T_1\) and (b) \(T_2\) in the ontological model defined over disjoint ontic states. Note that the resultant probability distribution from the convex combination is the same for (a) and (b) when the ontic states are converted to their non-disjoint counterparts and Eqs.~\ref{eq:nondisjointthreepairseq1}-\ref{eq:nondisjointthreepairseq3} are used. On the other hand, if this additional system of equations is ignored then the resultant probability distributions are different for \(T_1\) and \(T_2\). In fact, the two different resultant probability distributions correspond to the light and dark grey regions of the eight-state model's ontic space indicated in Fig.~\ref{fig:8stateonticspace}.}
  \label{fig:onticspaceconvexcombos}
\end{figure}

We can use the properties introduced in this section, along with the eight-state Edwards-Venn diagram, to show that the Grassmann WWM ontological model does not exhibit transformation contextuality under the ``blowtorch'' map. By using the tuples in \(\Lambda_6\) (the same as the set \(\bar g\) used in~\cite{Kocia17_2} and defined by Eq.~\ref{eq:probdistrogbar}) to organize the probability distributions of states \(\rho\), we can make use of the fact that their entries correspond to the coefficients in front of the monomials of a state's corresponding Weyl symbol, and thereby rely on Eq.~\ref{eq:identitytransformWeyl}-\ref{eq:HadtransformWeyl} to see how the states evolve under the Clifford gates \(X\), \(Y\), \(Z\) and the Hadamard \(H\). In this way, we see that under the \(T_1\) transformation,
\begin{eqnarray}
  &&(w_1,w_2,w_3,x_1,x_2,x_3) \\
  &\underset{T_1}{\rightarrow}& \frac{1}{4}[(w_1,w_2,w_3,x_1,x_2,x_3) \nonumber\\
  \label{eq:T1left}
  &&+ (x_1,x_2,w_3,w_1,w_2,x_3) \\
  &&+ (w_1,x_2,x_3,x_1,w_2,w_3) \nonumber\\
  &&+ (x_1,w_2,x_3,w_1,x_2,w_3) ] \nonumber\\
  \label{eq:T1right}
  &=& (0,0,0,0,0,0).
\end{eqnarray}
This is the probability distribution for \(\hat I/2\) and is illustrated in Fig.~\ref{fig:onticspaceconvexcombos}a. The final simplification exhibited in Eq.~\ref{eq:T1right} can be calculated in at least two ways: 
\begin{enumerate}
    \item Convert from disjoint (\(W_i\) and \(X_i\)) to non-disjoint (\(A_i\) and \(B_i\)) ontic states, employ Eqs.~\ref{eq:nondisjointthreepairseq1}-\ref{eq:nondisjointthreepairseq3} to simplify, and then convert back to disjoint states, or
    \item Find the Weyl symbol \(\rho' = T_1 \rho\) and then use Eq.~\ref{eq:proboflambdainrho} to obtain the probabilities \(P_{\rho'}(\lambda)\) that make up the entries of the resultant \(6\)-tuple \(\bar g\).
\end{enumerate}
These two methods are equivalent because, as discussed, by construction, Eqs.~\ref{eq:nondisjointthreepairseq1}-\ref{eq:nondisjointthreepairseq3} are a probability theory that captures the Weyl algebra.

Notice that it is not possible to obtain this solution without appealing to the ``coarse'' ontic states \(A_i\) and \(B_i\) either through method \(1\) or \(2\). Otherwise, a convex combination of four probability distributions from four different classes of solutions cannot be evaluated while satisfying the constraints given by Eqs.~\ref{eq:disjointthreepairseq1}-\ref{eq:disjointthreepairseq4}.


On the other hand, acting on line~\ref{eq:T1left} subsequently with the Hadamard gate to effect transformation \(T_2\) produces:
\begin{eqnarray}
  &&(w_1,w_2,w_3,x_1,x_2,x_3) \\
  &\underset{T_2}{\rightarrow}& \frac{1}{4}[(w_3,x_2,w_1,x_3,w_2,x_1) \nonumber\\
  \label{eq:T2left}
  &&+ (w_3,w_2,x_1,x_3,x_2,w_1) \\
  &&+ (x_3,w_2,w_1,w_3,x_2,x_1) \nonumber\\
  &&+ (x_3,x_2,x_1,w_3,w_2,w_1) ] \nonumber\\
  \label{eq:T2right}
  &=& (0,0,0,0,0,0).
\end{eqnarray}
Again, this is the probability distribution for \(\hat I/2\) and the final simplification exhibited in Eq.~\ref{eq:T2right} is the unique \(\bar g_\rho\) tuple for the final state. This can be seen in Fig.~\ref{fig:onticspaceconvexcombos}b. Notably, Fig.~\ref{fig:onticspaceconvexcombos} also shows that ignoring the additional constraints given by Eqs.~\ref{eq:nondisjointthreepairseq1}-\ref{eq:nondisjointthreepairseq3} leads to the inequivalent parity simplices in probability space that are found when \(T_1\) and \(T_2\) are implemented in the eight-state model. 

Therefore, the resultant probability distribution \((0,0,0,0,0,0)\) is attained no matter whether transformation \(T_1\) or \(T_2\) is taken in the Grassmann ontological model and so no transformation contextuality is present. The final solution is very similar to the one found in Section~\ref{sec:example}'s Eq.~\ref{eq:exampleconvexcombo} when using the ``coarse'' ontic states \(A\), \(B\) and \(C\). On the other hand, if the additional constraints are ignored and so a single convex set under the disjoint ontic states is assumed to exist, then different probability distributions are obtained under \(T_1\) and \(T_2\).

We further note that \emph{any} map consisting of convex combinations of states within the stabilizer subtheory will necessarily produce a unique probability distribution for every unique quantum state expected, since every Weyl symbol and one-qubit operator is bijectively represented by a probability distribution by the definition of \(\bar g\)~\cite{Kocia17_2} and we have shown that one-qubit convex combination is fully treated at order \(\hbar^0\) in Section~\ref{sec:grassmannHVT}. So this result of no transformation contextuality generalizes to all one-qubit maps within the one-qubit stabilizer subtheory.

We proceed to now show why the proof used by Lillystone \emph{et al}. explicitly excludes classical probability distributions defined over non-disjoint elements, such as those satisfying additional constraints like those given by Eq.~\ref{eq:nondisjointthreepairseq1}-\ref{eq:nondisjointthreepairseq3}. We then argue that the Grassmann WWM ontological model is proof that single qubit non-contextuality can be handled by ontological models with non-disjoint ontic states.

\section{Inequivalence Between Ontological Models with Disjoint Ontic States and Non-Disjoint Ontic States}
\label{sec:inequivalence}

Lillystone \emph{et al}. consider an arbitrary preparation non-contextual ontological model of a single qubit stabilizer subtheory. The WWM formalism for a single qubit is such a theory at
\(\hbar=0\). They then consider \(\Delta_p\) to be the support of the quantum state \(\rho\) in the ontological model,
\begin{equation}
  \Delta_p = \{\lambda | \mu_\rho(\lambda) > 0, \lambda \in \Lambda \}.
\end{equation}

The proof then proceeds to delete any state \(\lambda \in \Lambda\) such that \(P_{I/2}(\lambda) = 0\) and partition the remaining set into eight disjoint spanning sets. Since \(P_1(\lambda \setminus -\lambda) = 0\) for all \(\lambda \in \Lambda\), it follows by Eq.~\ref{eq:nondisjointthreepairseq1}-\ref{eq:nondisjointthreepairseq3} that \(P_1(\lambda \cap -\lambda) = \frac{1}{2}\) for all \(\lambda \in \Lambda\). Therefore, none of the eight Grassmann WWM ontic states given in Eq.~\ref{eq:Grassmannonticspace} are disqualified. 

Lillystone \emph{et al}. then proceed to produce a disjoint partition into eight sets. In particular, they rely on repeated application of the following feature of both \(\psi\)-ontic and \(\psi\)-epistemic ontologicla models: Given
\begin{equation}
  \label{eq:disjoint}
  \mu_\rho(\lambda) \mu_{\rho'}(\lambda) = 0 \hskip 5pt \forall \lambda\in \Lambda,
\end{equation}
this implies that
\begin{equation}
  \label{eq:disjointness}
  \mbox{supp} (\mu_\rho) \cap \mbox{supp}(\mu_{\rho'}) = \emptyset,
\end{equation}
if \(\rho \ne \rho'\)~\cite{Spekkens05}.

Since this is true for three pairs of basis states,~\cite{Lillystone18} argues that, given preparation non-contextuality, the ontic space can therefore be organized into \(2^3=8\) disjoint states. The argument is more clearly laid out in~\cite{Wallman12} and follows the reasoning that since six non-negative states have full support on only one unique basis element of one pair and the same partial support on all the other pairs, it must be possible to partition the space into eight disjoint sets.

For instance, in the eight-state model, the ontic states \(x=+\) and \(x=-\) are disjoint and so are \(y=+\) and \(y=-\). Hence, the ontic space can be partitioned into the four disjoint sets \(\{(x=+,y=+),(x=+,y=-),(x=-,y=+),(x=-,y=-)\}\). The partition into eight disjoint sets follows from then considering the disjoint sets \(z=+\) and \(z=-\).

In the Grassmann WWM qubit model, the ontic states \(\xi_r \xi_q \setminus \xi_q \xi_r\) and \(\xi_q \xi_r \setminus \xi_r \xi_q\) correspond to the eight-state model's \(x=+\) and \(x=-\) respectively, and \(\xi_p \xi_q \setminus \xi_q \xi_p\) and \(\xi_q \xi_p \setminus \xi_p \xi_q\) correspond to the eight-state model's \(y=+\) and \(y=-\) respectively. They can certainly be divided into the disjoint subsets by Lillystone \emph{et al}.'s argument and, along with the states \(xi_p \xi_r \setminus \xi_r \xi_p\) and \(\xi_r \xi_p \setminus \xi_p \xi_r\) that are analogous to the states \(z=+\) and \(z=-\) respectively, produce the eight disjoint ontic states given by Eq.~\ref{eq:Grassmannonticspace}.

But for the Grassmann WWM model, though Eq.~\ref{eq:disjointness} still holds (disjointness) and reexpressing ontic states in terms of disjoint ontic states produces bipartitions of the ontic space, it does not preserve convex combination of single-qubit state probability distributions. This is because the Weyl algebra over anti-commuting elements is equivalent to imposing additional constraints on top of Kolmogorov's axioms, as we have seen, which have more than one family of solutions when reexpressed in terms of disjoint ontic states and convex combinations between families of solutions is not preserved. This is true even though the model is preparation and measurement non-contextual for one qubit.

Lillystone \emph{et al}. complete the proof by directly relying on convex linearity to argue that there exists an implementation of \(T_1\) and \(T_2\) that has the same contextual implementation as theirs, when defined over the eight disjoint sets; they assume that the states in their convex sum fall into the same convex set when considered in terms of disjoint ontic states. They thus implicitly neglect the possibility of an additional set of constraints, commensurate with Kolmogorov's axioms, that does not result in a single convex set that contains all the probability distributions they consider when their ontic states are expressed as disjoint ontic states. Therefore, the Grassmann WWM ontological model lies outside the scope of their argument.

\section{Conclusion}
\label{sec:conc}

This paper answers the question of how to interpret the Grassmann WWM formalism in the framework of ontological models established by Leifer~\cite{Leifer14}. We show that the Grassmann WWM ontological model is \(\psi\)-epistemic, that it is most simply expressed in terms of overlapping or non-disjoint ontic states due to an additional set of constraints it must satisfy, that the probability distributions over these ontic states form a single convex set, but that probability distributions over disjoint ontic states do not.

If the additional set of constraints is ignored (i.e. explicitly not satisfied), convex combinations of probability distributions from different families of solutions produce different results when only one is expected. This is the origin of transformation contextuality under the ``blowtorch'' map in Lillystone \emph{et al}.'s ontological eight-state model over disjoint ontic states. We showed that transformation contextuality is not present in the Grassmann WWM classical probability theory at order \(\hbar^0\)---an ontological model defined over non-disjoint states with such an additional set of constraints. 

The Grassmann model offers a case where Lillystone \emph{et al}.'s proof---that preparation non-contextual qubit ontological models exhibit transformation contextuality in the one-qubit stabilizer subtheory---does not hold. Indeed, ontological models defined over non-disjoint ontic states appear to be able to treat single-qubit noncontextuality properly, and so do not exhibit transformation contextuality in the one-qubit stabilizer subtheory. We therefore contest Lillystone \emph{et al}.'s conclusion that ``the single-qubit stabilizer subtheory, a very simple subtheory of the smallest quantum system, exhibits generalized contextuality [and] demonstrates that generalized contextuality is so prevalent that even an essentially trivial quantum subtheory is classified as contextual, and therefore non-classical.''~\cite{Lillystone18}

In summary, we have shown that for the qubit stabilizer subtheory to be captured by a preparation, transformation and measurement non-contextual ontological theory, it must be handled in terms of non-disjoint ontic states, unlike the case for the odd-dimensional single-qudit stabilizer subtheory.

As a final point, one can ask more precisely why supplementation of Kolmogorov's axioms by an additional set of constraints does not seem to be present in the literature on ontological models so far. We point out that such an additional set of constraints can always be formulated for any ontological model after it is reexpressed in terms of non-disjoint ontic states. However, prior work has almost always considered ontological models where such an additional set of constraints is trivial because they are too weak; it only produces one family of solutions when the model is reexpressed in term of its original disjoint ontic states. And therefore it is natural that such constraints have not been discussed. Nevertheless, this ability to include an additional set of constraints, available due to the freedom provided by the sum rule (a consequence of \(\sigma\)-additivity), has always been there. In a way, this is an unused ``degree of freedom'' that has been hidden in plain sight all along, or at least since the introduction of the Grassmann algebra in physics.

\acknowledgments
Parts of this manuscript are a contribution of NIST, an agency of the US government, and are not subject to US copyright. P. L. acknowledges support from the National Science Foundation and Google LLC.

\bibliography{biblio}{}
\bibliographystyle{unsrt}

\end{document}